\documentstyle[aps,epsf,prb,multicol]{revtex}
\begin{document}
\draft
\title{Extrinsic levels, diffusion,  and
unusual incorporation mechanism of lithium in GaN} 
\author{Fabio Bernardini$^{(1)}$ and
 Vincenzo Fiorentini$^{(1,2)}$} 
\address{(1)\, Istituto Nazionale per la Fisica della Materia and
Dipartimento di Fisica, 
Universit\`a di Cagliari, Italy\\
(2)\, Walter Schottky Institut, Technische Universit\"a{}t M\"u{}nchen, 
 Germany}
\date{\today}

\maketitle 
\begin{abstract}
Results of a first-principles study of the Li impurity in GaN are
presented. We find Li  is a channel interstitial,  with an onset for
diffusion at  T$\sim$ 600 K. Above this temperature, Li can transform
to a Ga-substitutional acceptor by exothermic recombination with Ga
vacancies. This process implies capture of at least one electron;
therefore Li acts as an  electron sink.
Li$_{\rm Ga}$ is stable
again interstitialcy, and has a shallow   first ionization levels of
0.16 eV, and  second ionization at 0.63 eV. Lattice locations and
their temperature dependence are in close  agreement with  recent
experiments. 
\end{abstract}
  % impurity concentration 

\begin{multicols}{2}
Doping III-V nitrides $p-$type is a conceptually intriguing 
and practically relevant problem. 
 The current practical recipe  for GaN $p-$doping is  based on the Mg
acceptor, but the search for alternative dopants is still ongoing.
Useful informations on the mechanism involved in Mg
incorporation,\cite{neu.apl} and on possible  alternative
acceptors,\cite{noi.apl} has come from  theoretical
investigations. While Group-II dopants have been treated in detail,
\cite{neu.apl,noi.apl,noi.icps,neu.fkp} little attention has been
devoted so far to potential  Group-I double acceptors.\cite{jap.vdw}

In this Communication we discuss the physics of 
 Li as an impurity in GaN,
building on direct calculations of energetics, equilibrium geometries, and
diffusion paths. Li is found to be preferentially 
incorporated in GaN as an interstitial. Since its chemical incorporation is 
 costly, incorporation should be effected by ionic implantation. 
Diffusion of Li in GaN is activated, with a threshold temperature of
about 600 K, and  occurs in three dimensions with a slight preference for
motion within the basal plane.  Li  recombination  with Ga
vacancies   is a  highly exothermic, barrier-less reaction. Hence, if
diffusion is activated,  Li is efficiently  incorporated substitutionally into
 any available Ga vacancy. In the recombination process,  capture  of one
or two electrons  occurs for almost all values of
 the Fermi level E$_{\rm F}$. The substitutional double acceptor 
Li$_{\rm Ga}$ thus realized
is highly stable again interstitialcy (as the Ga vacancy-Li
interstitial pair is highly disfavored with respect to Li$_{\rm Ga}$)
and, most importantly, it exhibits a thermal first ionization energy
of 0.16 eV, smaller than the
ionization energy  of Mg$_{\rm Ga}$, and a second ionization level 
at  0.63 eV. Because of the strong dependence on precursor
defects (the Ga vacancy), the role of Li as an acceptor or
compensator will be significant
only if its precursors are the dominant native defects of the system.
Of course, this   limits its practical relevance for GaN  doping.

We used  local density-functional theory \cite{dft}
 ultrasoft-pseudopotential\cite{uspp} plane-wave
 calculations  of  total energies
 and forces   in   GaN wurtzite supercells (typically encompassing
32 atoms), and with plane wave cutoff of 25  Ry,  to predict from first
principles the total energies and equilibrium geometries  relevant to 
our problem. All atomic coordinates are fully relaxed until 
residual forces are below 0.01 eV/\AA. Further technicalities and
the  formalism to
calculate defect  formation energies have been reported previously in 
detail (see e.g. Refs. \onlinecite{neu.apl} and \onlinecite{noi.apl}).
One point  worth mentioning here is the  solubility limit being
 metallic bulk Li, rather than Li$_3$N as  could be expected.

\begin{figure}
\narrowtext
\epsfclipon
\epsfxsize=8cm
\centerline{\epsffile{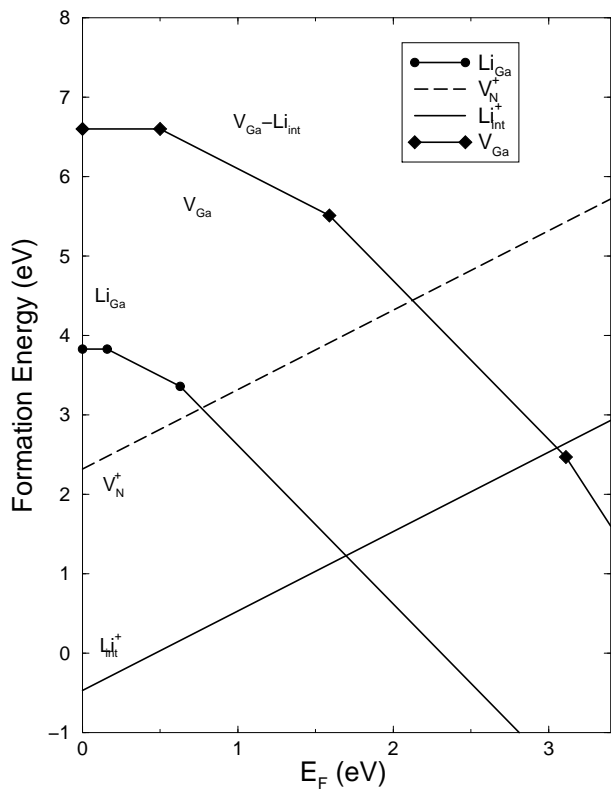}}
\caption{Formation energies of Li and native defects
discussed in the text, in N-rich conditions.}\label{fig0}
\end{figure}

The formation energies of relevant defects and impurity configurations 
are summarized for reference in Figure \ref{fig0}. We begin  by
examining the substitutional  Li$_{\rm Ga}$. In this configuration  Li
behaves as a double acceptor: we  determined directly its  
first and second thermal ionization energies, that are 0.16 eV and
0.63 eV, respectively, above the valence band.\cite{nota2} This gives
us additional motivation for this study, since the first ionization
energy is comparable than that of Mg$_{\rm Ga}$; the second
ionization is fourfold the first as expected from hydrogenic 
models.\cite{noi.icps,neu.fkp}

Since the chemical formation energy of Li$_{\rm Ga}$
is rather high (3.83 eV in the neutral state in N-rich conditions), plain 
chemical solubility does  not seem an easily viable incorporation  pathway  
for substitutional Li in GaN in moderately $p-$type conditions.
 Ion implantation\cite{egw} could
 possibly surrogate chemical reaction. It is then natural to investigate 
 precursors
to substitutional incorporation, selected among native defects and
Li complexes therewith. Specifically, we studied the Ga vacancy and
 a selection of  Li interstitials and  V$_{\rm Ga}$--Li$_{\rm int}$
pairs. We also mapped out the diffusion path and  estimated the activation
barriers for Li$_{\rm int}$ motion along high symmetry directions  in 
undefected wurtzite GaN, and towards a neighboring Ga vacancy in 
cation-vacancy  defected GaN.

The results for V$_{\rm Ga}$ and V$_{\rm N}$
are quite compatible with previous 
calculations.\cite{neu.fkp} The cation vacancy, in particular,
 is a deep triple acceptor, with
formation energy going from fairly high in $p-$type conditions to 
very low in $n-$type. Local relaxations are in agreement with
previously reported data.

As for Li$_{\rm int}$ in  undefected GaN, we looked for stable sites in the
open channels of the wurtzite structure and in the adjacent
trigonal cages.
 In all configurations the stable charge state is +1.
 We found that the marked preference for antibonding or 
bond-center configurations  exhibited  \cite{neu.prl,noi.mrs}
by hydrogen in GaN
is not shared at all by Li. None of the various sites suggested
by a naive analogy with H is found to be stable.
  The lowest-energy configuration of Li$_{\rm int}$  is at the center
 of the  open wurtzite channel, and has C$_{3v}$ symmetry. The
 site can be viewed as delimited at the top and bottom by 
three-fold asymmetric boat-shaped Ga-N rings, and laterally by
symmetric boat-shaped three-fold rings.
The coarse-grained picture is that Li simply tries to maximize
 its distance from all other atoms, but on closer inspection
a slight preference is revealed towards N atoms. 
In particular, Li$_{\rm int}$  sits at 3.61 bohr from 
 the N atoms protruding from the upper ring, 
and at 4.33 bohr from the  Ga atoms in the lower ring.
The equilibrium sites  along the $c$ axis are
0.23 $c$ and 0.73 $c$, which compare well with the
measured \cite{egw} values of 0.25 $c$ and 0.75 $c$
[the implicit convention here is that Ga atoms have $z$-coordinates
zero and 0.5 $c$, and N atoms have 0.376 $c$ and 0.876 $c$ ]. 
Inside the trigonal cage, we only find a marginally  stable
site near the cage center,
at about 1.3 eV higher than the channel site,
and with a confinement barrier of only 0.1 eV.
Again, Li prefers to sit closer
to N atoms (Li-N distance $\sim$ 3.18 bohr) than to
Ga atoms (Li-Ga distance $\sim$ 3.59 bohr).

These results can be largely understood
with a simple Madelung-like electrostatic model, whereby
we calculate the electrostatic energy of 
channel-interstitial Li as a function of
its coordinate along the $c$ axis, assuming it interacts
with point-charge ions placed on a wurtzite lattice. 
N and Ga are assigned their dynamical Born 
charge \cite{noi.piezo} of $\pm$ 2.7, while
 Li is assumed to have a dynamical charge of 0.9.
The  equilibrium site is found  at 0.31 $c$, close to 0.23 $c$ and
0.25 $c$ given by direct calculation 
and experiment, respectively; the barrier for 
displacement along $c$ is 3 eV, which  compares well 
with 2.7 eV from direct calculation for the unrelaxed lattice.
 This shows that basically Li sits at
the site dictated by the minimization of the electrostatic interactions with
the ionic lattice. The sizable deviation in the equilibrium coordinate
is due to the large size of the N ions, assumed to be pointlike in
the model.

Starting from these sites, we investigated the diffusion paths for Li
in GaN.The barrier for diffusion between equilibrium sites  along 
the (0001) direction is 1.55 eV, and the 
vibrational frequency at the equilibrium site along (0001) is
13 THz. When 
the diffusion path from channel to channel across a cage is considered,
the  maximum barrier  is found to occur near the cage center, at 1.4
eV above the channel site. At temperatures at which this
diffusion path is activated, the previously mentioned
marginally  stable pocket near  the cage center is of course irrelevant,
 given its confinement barrier  of only $\sim$ 0.1 eV. 
 The vibrational frequency at the channel equilibrium site 
along the  cross-cage diffusion direction is 9 THz. 
The  effective activation temperature for Li diffusion can be 
estimated from
the above data to be about 600 K.
At the activation temperature, cross-cage motion, i.e. motion 
approximately within the basal plane, is more probable than 
motion along the $c$ axis by a factor of 10 on account of its lower 
barrier.

\begin{figure}
\narrowtext
\epsfclipon
\epsfxsize=5cm
\centerline{\epsffile{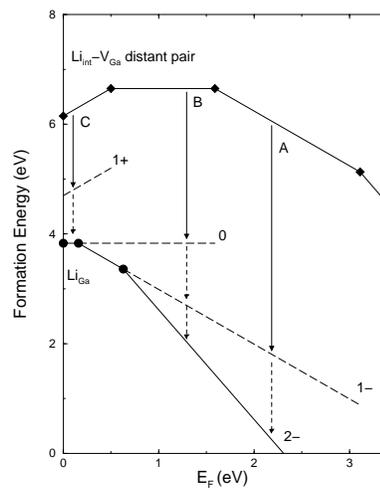}}
\caption{Formation energies of Li$_{\rm Ga}$ and
the Li$_{\rm int}$--V$_{\rm Ga}$  pair.
Vertical lines schematize possible recombination 
processes. Dashed lines indicate metastable charge states of
Li$_{\rm Ga}$.}\label{fig1}
\end{figure}

We now consider  the
propensity of Li to recombine into Ga vacancies, which we assume
henceforth to be {\it already present} in the system. 
In Figure \ref{fig1} we compare
 a distant  Li$^+$-V$_{\rm Ga}$ pair\cite{nota} with
 substitutional Li$_{\rm Ga}$;
the latter is always vastly favored energetically
(by over 5 eV for a semi-insulating crystal,
i.e.  midgap Fermi level), so Li-vacancy
recombination is highly favorable.
This recombination process involves
 several  steps. First, Li diffuses across the crystal
towards the neighborhood of a
vacant Ga site; as discussed
 previously, this requires a temperature above about 600
 K.  Second, Li recombines with the vacancy: upon direct calculation,
we find that  Li drops into V$_{\rm Ga}$ from the wurtzite channel
  {\it  without any activation barrier},
irrespective of the vacancy charge state.
 Third, the substitutional Li thus produced 
assumes a  charge state consistent with the Fermi
 energy.

As to the latter step we note that,
 on account of the charge state of the precursor
 Ga vacancy and depending on the Fermi level,
the metastable V$_{\rm Ga}$-Li$_{\rm int}$ distant pair 
 exists in charge states from +1 to --2. Upon recombination with the
 vacancy, Li$_{\rm Ga}$ may   therefore need to capture electrons
 to reach its equilibrium  charge  state  (0 to  --2,
but in fact equal to the latter for most of the accessible E$_{\rm F}$
 range). 
The level diagram  in Fig. \ref{fig1} shows indeed that either one or two
 electrons will be captured over practically all  the Fermi  level
 range. For example,  in process A  the pair is singly negative
(Li  recombines with V$_{\rm Ga}^{-2}$ ), hence Li captures one
 electron to reach its equilibrium charge state of --2.
 In process  B,  Li recombines with  V$_{\rm Ga}^{-1}$
 (neutral pair) and capture of two electrons ensues.  Processes in
 $p$-type conditions, such as C in Fig. \ref{fig1}, are
 complicated by  the acceptor levels of Li$_{\rm Ga}$, but  all imply
 the capture of either one or two electrons.  

In short, the reason is that all charge states of the Li-vacancy pair 
 overlap in energy only with {\it more negative} charge
states of Li$_{\rm Ga}$. The only exception is the doubly negative
 Li-vacancy pair (Li recombination with  triply negative Ga
 vacancies): the latter,  however, only  occurs in extreme  $n$-type
conditions, where in fact the direct  chemical 
incorporation of Li becomes favorable. 
For all other  Fermi levels, Li acts as an electron sink,
 i.e.  as an effective  acceptor, or more properly, compensator.  
 Fig. \ref{fig1} also shows that electron capture is more
likely at lower  Fermi levels, {\it i.e.} towards $p$-type conditions:
 in particular, if Li$_{\rm Ga}$ were always --2, the number of captured
electrons  would change from 0 through 3 as E$_{\rm F}$ moves
downwards across  the gap. This is quite clearly the opposite of what
happens to normal  ``chemical'' acceptors. Unfortunately,
 the mechanism is limited by the availability of Ga vacancies as
 precursors; any Li  not paired to a Ga  
vacancy will  remain interstitial.

Let us  summarize the picture we arrived at. Isolated  Li
 in GaN  sits in the wurtzite channels. It will be very strongly
 attracted towards any neighboring Ga vacancy, and it will recombine
 with it to give rise to  substitutional Li$_{\rm Ga}$. 
 To reach  a vacancy, Li must
 diffuse into GaN;  diffusion  sets in at
  approximately 600 K (with a preference
for the basal plane near the onset). Once in
the neighborhood of the vacancy, Li  recombines with the vacancy
  without further  energetic barrier  hindering the process.
 Given the  gain upon  recombination of Li  into the 
vacancy,   the reverse  process  Li$_{\rm Ga} \rightarrow$  Li$_{\rm
 int}$ + V$_{\rm Ga}$ is competely ruled out, {\it i.e.} 
Li$_{\rm Ga}$ is stable again interstitialcy.  At the end of the
 recombination  transient, the  Li$_{\rm Ga}$  concentration will  
 equal  that of the pre-existing Ga vacancies (which will have 
vanished),  and the  electron density will have been reduced by 
one to two times the  smaller of the two concentrations of Li$_{\rm int}$
 and V$_{\rm Ga}$, depending on  the initial  Fermi level. 
Clearly, compensation of $n$-type carriers will effectively
occur if Li is incorporated in GaN in its +1 ionized state, as in ion 
implantation. Any  Li population in excess of the  
Ga vacancy concentration will remain interstitial.

We close our investigation making
 contact with the only piece of  experimental evidence on
 Li in GaN, a study of decay emission from  implanted
radioactive Li isotopes. \cite{egw} At room
temperature, channeling  $\gamma-$ray emission   is
observed, showing that Li is indeed interstitial; as mentioned
 earlier on in this paper, our calculated locations agree closely
with those  deduced from angle-dependent  channeling.
After annealing at 700 K, the $\gamma$ emission from Li decay 
 is shadowed, indicating that Li  has either  moved into a trigonal
 cage, or to a substitutional site. On the basis  of our results,
 the natural explanation  is Li-Ga vacancy  recombination. 
While shadowing may occur  also if Li would  remain confined in a
 trigonal cage, our calculated diffusion barriers  show that this is
 not possible, as the cage center is not a stable site at the
 relevant  temperature. Our  estimate of  600 K  for the onset of
 diffusion  (and consequent  recombination with Ga vacancies)    is in
 rather good agreement with  experiment, given our neglect of  
 entropic contributions to the free energy diffusion barrier.

In summary, we have shown by  first-principles calculations
that Li can be easily incorporated  as an interstitial in GaN, 
and that it will efficiently recombine with cation vacancies to yield
Li$_{\rm Ga}$. The resulting substitutional double acceptor is predicted 
to have first and second thermal ionization energies at 0.16 eV
and 0.63 eV above the valence band, energies which are comparable or
smaller to those of acceptors commonly used in GaN. Due to
limitations due to the availability of
its precursor V$_{\rm Ga}$, Li is expected to   
function only as compensator of  $n$-type carriers.

We acknowledge dedicated 
 funding from INFM through Iniziativa Calcolo Parallelo and the 
 Computational Semiconductor Physics Network of Section E.
  VF was supported by  the Alexander von Humboldt-Stiftung  during his 
stay at WSI.

\end{multicols}

\begin{references}
\bibitem{neu.apl}
J. Neugebauer and C. G. van de Walle,  
Appl. Phys. Lett. {\bf 68}, 1829 (1996);
\bibitem{noi.apl}
F. Bernardini, V. Fiorentini, and A. Bosin,
Appl. Phys. Lett. {\bf 70}, 2990 (1997).

\bibitem{noi.icps}
V. Fiorentini, F. Bernardini, A. Bosin, and D. Vanderbilt,
in {\it The Physics of Semiconductors}
M. Scheffler and R. Zimmermann eds. (World Scientific, 1996), p. 2877.
F. Bernardini, V. Fiorentini, and R. M. Nieminen,
 {\it ibid}, p. 2881,

\bibitem{neu.fkp}
J. Neugebauer and C. G. van de Walle,  
 Fest\-k\"or\-per\-pro\-bl\-eme/Advances  in
Solid State Physics, vol. 35, R. Helbig ed. (Vieweg, Braunsch\-weig
1996), p.25.

\bibitem{jap.vdw}
J. Neugebauer and C. G. van de Walle, 
J. Appl. Phys. {\bf 85}, 3003 (1998).

\bibitem{dft}
 R. Dreizler and E. Gross, {\it Density functional theory}, (Springer,
Berlin, 1990).  

\bibitem{uspp}
D. Vanderbilt, 
Phys. Rev. B {\bf 41}, 7892 (1990).

\bibitem{nota2}
The thermal first ionization energy $\epsilon[0/-]$, corresponding
 to hole release from  the  acceptor state,
is the formation-energy difference of charge
states $Q$=--1 and $Q$=0 at $\mu_e$=0.
The second ionization energy is the same as above for charge
states $Q$=--2 and $Q$=--1.

\bibitem{neu.prl}
 J. Neugebauer and C. G. van de Walle,  
Phys. Rev. Lett.
{\bf 75}, 4452 (1995).



\bibitem{egw}
M. Dalmer {\it et al.}, J. Appl. Phys. {\bf 84}, 3085 (1998).
\bibitem{noi.mrs}
 A. Bosin, V. Fiorentini, and D.  Vanderbilt, in {\it Gallium Nitride
and related compounds},  R. D. Dupuis, J. A. Edmond, F. Ponce, and
S. Nakamura eds., MRS Proceedings {\bf 395}, 503 (1996).

\bibitem{noi.piezo}
F. Bernardini, V. Fiorentini, and D. Vanderbilt, 
Phys. Rev. B {\bf 56}, R10024 (1997).

\bibitem{nota}
The formation energy of this pair
is the sum of the formation energies of Li$_{\rm int}$ and
 V$_{\rm Ga}$, calculated individually in separate simulation
cells, {\it i.e.} for non-interacting defects. Conversely,
the 
substitutional is the strongly interacting limit for
this pair.
\end{references}
\end{document}